\documentstyle[aps,epsfig,floats]{revtex}
\def\lsi{\raise0.3ex\hbox{$<$\kern-0.75em\raise-1.1ex\hbox{$\sim$}}}
\def\gsi{\raise0.3ex\hbox{$>$\kern-0.75em\raise-1.1ex\hbox{$\sim$}}}
\newcommand{\lsim}{\mathop{\lsi}}

\begin{document}
\draft
\twocolumn[\hsize\textwidth\columnwidth\hsize\csname
@twocolumnfalse\endcsname
\title{Determination of absolute neutrino masses from Z-bursts}
\author{
Z.~Fodor$^{a,b}$, S.D.~Katz$^b$ and A. Ringwald$^a$}
\address{$^a$Deutsches Elektronen-Synchrotron DESY, Notkestr. 85, D-22607,
Hamburg, Germany\\
$^b$Institute for Theoretical Physics, E\"otv\"os University, P\'azm\'any
1, H-1117 Budapest, Hungary}

\date{\today}
\maketitle

\vspace*{-4.0cm}
\noindent
\hfill \mbox{ITP-Budapest 567\ \ \ \ DESY 01-048}
\vspace*{3.8cm}

\begin{abstract} \noindent
Ultrahigh energy neutrinos (UHE$\nu$) scatter on 
relic neutrinos (R$\nu$) producing Z bosons, which can decay hadronically 
producing protons (Z-burst). We 
compare the  predicted proton spectrum with the observed 
ultrahigh energy cosmic ray (UHECR) spectrum and determine the  
mass of the heaviest R$\nu$ via a maximum
likelihood analysis. 
Our 
prediction depends on the origin of the power-like part
of the UHECR spectrum: 
$m_\nu=2.75^{+1.28}_{-0.97}$ eV 
for Galactic halo and 
$0.26^{+0.20}_{-0.14}$~eV for extragalactic (EG) origin. 
The necessary UHE$\nu$ 
flux 
should be detected in the near future. 
\end{abstract}

\vspace*{0.2cm}
\vskip1.3pc]

{\it I. Introduction.\,---}
The interaction of protons ($p$) with photons ($\gamma$) of the cosmic microwave
background radiation (CMBR) predicts a sharp drop in the cosmic ray flux 
above the Greisen-Zatsepin-Kuzmin (GZK) cutoff around 
$4\cdot 10^{19}$~eV~\cite{GZK66}. The
available data show no such drop. About 20 events above $10^{20}$~eV
were observed by experiments such as 
AGASA~\cite{AGASA}, Fly's Eye~\cite{FLY}, Haverah Park~\cite{HAVERAH},
Yakutsk~\cite{YAKUTSK}, and HiRes~\cite{HIRES}. 
The attenuation length of protons above the GZK cutoff is $\approx$~50~Mpc; 
but no obvious astrophysical source candidate is known within this 
distance. No conventional explanation for the observed UHECR spectrum
is known~\cite{BS00}.

Already in the early 80's there were discussions that the UHE$\nu$ spectrum
could have absorption dips at energies around
$E_{\nu_i}^{\rm res} = M_Z^2/(2\,m_{\nu_i}) = 4.2\cdot 10^{21}$ 
(1 eV/$m_{\nu_i}$) eV 
due to resonant annihilation with the R$\nu$s, 
predicted by the hot Big Bang cosmology, 
into Z bosons of mass $M_Z$~\cite{W82,Y97}. 
Recently it was realized that the same annihilation mechanism gives 
a possible solution to the GZK problem~\cite{FMS99}. 
It was argued that the UHECRs above the GZK cutoff are 
mainly from Z-bursts taking place within the GZK zone of $\approx$ 50 Mpc. 

This hypothesis was discussed in several papers~\cite{Y99,W98,GK99,Gelmini:2000bn,PW01,FGDTK01}. 
In Ref.~\cite{Y99}, particle 
spectra were determined numerically for case studies which supported the 
Z-burst scenario. The required UHE$\nu$ fluxes for different spectral
indices were calculated in Ref.~\cite{W98}, too. 
The effect of possible lepton asymmetries was studied in Ref.~\cite{GK99}.
In Ref.~\cite{PW01}, the analysis of the Z-burst mechanism was advocated as 
one of the few possibilities for an absolute $\nu$ mass determination and 
its potential compared to others like e.\,g. the $\beta$ decay endpoint 
spectrum and the $\nu$-less $\beta\beta$ decay. 

There is now ra\-ther convincing evidence that $\nu$s have nonzero masses 
(cf.~\cite{PDG}). This evidence comes from $\nu$ oscillation measurements 
with typical mass splittings $\sqrt{\delta m^2} \sim 10^{-5}\div 0.4$ eV.
Neutrinos in this mass range are important 
cosmologically since they represent a 
non-negligible contribution to dark matter (DM) which imposes upper limits
on $\nu$ masses~\cite{MS74}. Hydrodynamic
simulations with massive $\nu$s and including recent observational
measurements and cosmological constraints give~\cite{CHD99} 
$\sum_i m_{\nu_i} \lsim 2.4 \cdot (\Omega_M/0.17-1)$ eV, if 
the matter content of the universe $\Omega_M$ is assumed to be between
0.2 and 0.5, as favoured by recent measurements (cf.~\cite{PDG}).

{\it II. Z-burst spectrum and UHECR data.\,---}
Our comparison of the Z-burst scenario with the observed UHE\-CR spectrum
is done in four steps. First we determine the probability
of Z production as a function of the distance from earth.
In the second step we exploit collider experiments to derive 
the energy distribution of the produced protons in the lab system.
The third ingredient is the propagation of the protons, i.\,e. the 
determination of their energy loss due to pion and $e^+e^-$ production 
through scattering on the CMBR and due to their redshift. The last
step is the comparison of the predicted and observed spectrum and 
the extraction of the mass of the R$\nu$ and the necessary UHE$\nu$ flux.   

For a given neutrino type $i$ the probability of Z-bursts 
at some distance $r$ is proportional to 
the number density $n_{\nu_i}(r)$ of the R$\nu$s and 
to the flux $F_{\nu_i}(E_{\nu_i},r)$ of 
the UHE$\nu$s at energy $E_{\nu_i}\approx E_{\nu_i}^{\rm res}$. 
The density distribution of R$\nu$s as hot DM follows the total mass 
distribution; however, it is expected to be less clustered.   
This is the reason why we, 
similarly to Ref.~\cite{Gelmini:2000bn} but in distinction to 
practically all previous
authors~\cite{FMS99,Y99,W98}, do not follow the 
assumption of having 
a relative overdensity of 
$f_\nu = 10^2\div 10^4$ in our neighbourhood. 
For distances below 100 Mpc we varied the shape of the $n_{\nu_i}(r)$ 
distribution between the homogeneous case and that of $m_{{\rm tot}}(r)$, 
the total mass distribution obtained from peculiar velocity 
measurements~\cite{peculiar}. 
Our results are rather insensitive to these 
variations. Their effect is included in our error bars. For scales
larger than 100 Mpc the R$\nu$ density is given by the Big Bang cosmology, 
$n_{\nu_i}=56\cdot (1+z)^3$ cm$^{-3}$. In our analysis we go up to distances 
of redshift $z = 2$ (cf. \cite{W95}). We include uncertainties  
of the expansion rate (see e.\,g. Sect. 2 of ~\cite{PDG}). 
The UHE$\nu$ flux is assumed to have the form 
$F_{\nu_i}(E_{\nu_i},r)=F_{\nu_i}(E_{\nu_i},0)\,(1+z)^\alpha$, where
$\alpha$ characterizes the source evolution (see also~\cite{Y97,Y99}).
Independently of the production
mechanism, $\nu$ oscillations result in a uniform $F_{\nu_i}$ mixture for the 
different types $i$. 

The Z-burst scenario is based on Z decays. 
At LEP and SLC millions of Z bosons were produced and their
decay analyzed with extreme high accuracy. $69.89\%$ of the Z decays 
are hadronic and the $p+{\bar p}$ multiplicity is 
$\langle N_p\rangle = 1.04\pm 0.04$ in the 
hadronic channel~\cite{PDG}. The neutron multiplicity, which we included in 
our analysis, is $\approx 4\%$ smaller than the proton's~\cite{Biebel01}.  
We combined 
existing published and some improved unpublished data on the momentum 
distribution 
${\cal P}\,(x=p_{\rm proton}/p_{\rm beam})$ of protons in Z 
decays~\cite{Zdecay}. 
Due to the large statistics, the uncertainties 
related to Z decay are negligible.

In the CM system of the Z production the
angular distribution of the hadrons is determined by the
spin $1/2$ of the primary quarks and thus proportional to 
$1+w^2=1+\cos^2\theta$ (here $\theta$ is the angle between the
incoming neutrinos and the outgoing hadrons (cf.~\cite{S95})).  
The energy distribution ${\cal Q}(E_p)$ of the produced protons with energy 
$E_p$ is finally obtained after a Lorentz transformation from the CM system
to the lab system,
\begin{eqnarray}
&&{\cal Q}(E_p)=
\frac{2}{E_\nu}\ 
\sum_{+,-} {3 \over 8} \int_{-1}^1 {\rm d}w\, (1+w^2)  \\
&&{1 \over 1-w^2}
\left|
{{\pm y-w\sqrt{y^2-(1-w^2)(2m_p/M_Z)^2}}
\over 
\sqrt{y^2-(1-w^2)(2m_p/M_Z)^2}}
\right|
\nonumber\\
&&{\cal P}\left([
-
wy \pm \sqrt{y^2-(1-w^2)(2m_p/M_Z)^2}]
/(1-w^2)\right), \nonumber
\end{eqnarray}
where $m_p$ is the $p$ mass and $y=2E_p/E_\nu$.  

Particles of EG origin
and energies above $\approx 4\cdot 10^{19}$ eV lose a large fraction
of their energies~\cite{GZK66}. 
This can be 
described by the
function $P(r,E_p,E)$, the probability
that a proton created at a distance $r$ with energy $E_p$ arrives
at Earth above the threshold energy $E$~\cite{BW99}. 
It has been calculated for a wide range
of parameters in Ref.~\cite{FK00},  
and the respective data are available at www.desy.de/\~{}uhecr. 
Note, that the energy attenuation length of $\gamma$s
are longer than that of protons roughly by a factor of 10 at superhigh
energies such as $10^{21}$ eV. 
The detailed study of the boosted Z-decay (data from Ref.~\cite{Zdecay}) results  in 
$\gamma$s of energy 
below  $10^{19}$ (1 eV/$m_{\nu_i}$) eV, 
where their attenuation length 
is much smaller,  
for strong enough radio background.  Thus,
their contribution 
to the UHECR spectrum is far less relevant
than that of the protons.

The Z-burst contribution to the UHECR spectrum, for 
degenerate $\nu$ masses ($m_\nu\approx m_{\nu_i}$), 
is given by
\begin{eqnarray}\label{nu-flux}
&&j(E,m_\nu)=I\cdot F_Z^{-1} \cdot 
\int_0^\infty {\rm d}E_p \int_0^{R_0} {\rm d}r 
\int_0^\infty {\rm d} \epsilon
\\ \nonumber
&&\sum_{i} F_{\nu_i}(E_{\nu_i},r)\sigma( \epsilon)n_{\nu_i}(r)
{\cal Q}(E_p)
\left(-\partial P(r,E_p,E) / \partial E\right),
\end{eqnarray}
where $I \approx 8\,\cdot 10^{16}$ m$^2\cdot$\,s\,$\cdot$\,sr is the 
total exposure
(estimated from the highest energy
events and the corresponding fluxes),  
$R_0$ is the distance at $z=2$, and  
$\sigma(\epsilon)$ is the Z production
cross section at CM energy $\epsilon = (2\,m_\nu\,E_{\nu_i})^{1/2}$.  
The normalization factor $F_Z$ is proportional to the
sum of the $\nu$ fluxes at CM energy $M_Z$.   

We compare the spectrum~(\ref{nu-flux}) with
the observed one and give the value of $m_\nu$ based on a maximum
likelihood analysis. In the Z-burst scenario a small R$\nu$ mass needs large 
$E_\nu^{\rm res}$ in order to produce a Z. Large $E_\nu^{\rm res}$ results in 
a large Lorentz boost, thus large $E_p$. In this way the
detected $E$ determines the mass of the R$\nu$.  

Our analysis includes the published and the unpublished
(from the www pages of the experiments on 17/03/01) UHECR data
of~\cite{AGASA,FLY,HAVERAH,HIRES}. Due to normalization difficulties
we did not use the Yakutsk~\cite{YAKUTSK} results. 

\begin{figure}[t]
\begin{center}
\epsfig{file=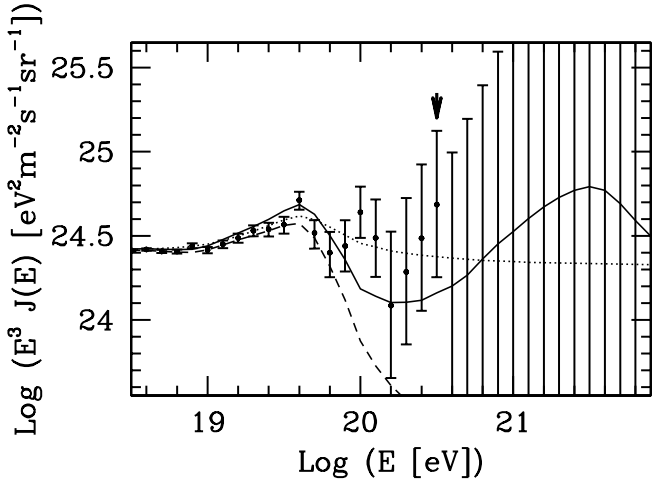,bbllx=20pt,bblly=225pt,bburx=570pt,bbury=608pt,%
width=8.65cm}
\caption{\label{fit_nu}
{The available UHECR data with their error bars
and the best fits from Z-bursts.
Note that there are no events above $3 \times 10^{20}$~eV
(shown by an arrow). 
The dotted line shows the best fit for the ``halo''-case. The bump around
$4\cdot 10^{19}$~eV is due to the Z-burst protons, whereas the almost
horizontal contribution is the first, power-law-like 
term of Eq.~(\ref{flux}). The solid line shows the ``extragalactic''-case.
The first bump at $4\cdot 10^{19}$~eV represents protons produced at
high energies and accumulated just above the GZK cutoff due to their energy
losses. The bump at $3\cdot 10^{21}$~eV is a remnant of the Z-burst
energy. The dashed line shows the contribution of the power-law-like
spectrum with the GZK effect included. The predicted fall-off for this
term around $4\cdot 10^{19}$~eV can be observed. 
The attenuation of the Z-burst component appears to be weaker on account of 
the narrowness of the injected proton spectrum and the fact that the
observed post-GZK protons are produced within the GZK zone. 
}}
\end{center}
\end{figure}

Since the Z-burst scenario results
in a quite small flux for lower energies, the ``ankle''  
is used as a lower end for the UHECR spectrum:
$\log (E_{\rm min}/\mbox{eV})=18.5$. Our results are
insensitive to the definition of the upper end (the flux is
extremely small there) for which we choose $\log (E_{\rm max}/\mbox{eV})=26$.
As usual, we divided each logarithmic unit into ten bins. The
integrated flux
gives the total number of events in a bin. The uncertainties of the
measured energies are about 30\% which is one bin. Using a Monte-Carlo method
we included this uncertainty in the final error estimates.
For the degenerate case, the predicted number of events in a bin is given by
\begin{equation}\label{flux}
N(i)=\int_{E_i}^{E_{i+1}}
{\rm d}E
\left[A \cdot E^{-\beta}+F_Z\cdot j(E,m_\nu)\right],
\end{equation}
where $E_i$ is the lower bound of the $i^{\rm th}$ energy bin. The first
term is the usual power law,  
which describes the data 
well for smaller energies~\cite{AGASA}. For this term we
will study two possibilities. In the first case we assume
that the power part is produced in our galaxy. Thus no
GZK effect should be included for it (``halo''). In the second -- in
some sense more realistic -- case 
we assume that the protons come from uniformly distributed, EG sources
and suffer from the GZK cutoff (``EG''). In this case the simple
power-law-like term will be modified and falls off around $4\cdot 10^{19}$~eV
(see later Fig.~\ref{fit_nu}). The second term of the
flux in Eq.~(\ref{flux}) corresponds 
to the spectrum of the Z-bursts, Eq.~(\ref{nu-flux}). 
A and $F_Z$ are normalization factors.

The expectation value for the number of events in a bin is given
by Eq.~(\ref{flux}) and it is Poisson distributed. To
determine the most probable value for $m_\nu$ we used the maximum likelihood
method and minimized~\cite{FK01} the $\chi^2(\beta,A,F_Z,m_\nu)$ for
Poisson distributed data~\cite{PDG}, 
\begin{equation} \label{chi}
\chi^2=\sum_{i=18.5}^{26.0}
2\left[ N(i)-N_{\rm o}(i)+N_{\rm o}(i)
\ln\left( N_{\rm o}(i)/N(i)\right) \right],
\end{equation}
where $N_{\rm o}(i)$ is the total number of observed events in the $i^{\rm th}$
bin. In our fitting procedure we have four parameters: $\beta,A,F_Z$ and 
$m_\nu$. The minimum of the $\chi^2(\beta,A,F_Z,m_\nu)$ function is 
$\chi^2_{\rm min}$
at $m_{\nu\, {\rm min}}$,
the most probable value for the mass. 
The 1\,$\sigma$ confidence
interval for $m_\nu$ is given by
$\chi^2(\beta',A',F_Z',m_\nu)=\chi^2_{\rm min}$+1. $\beta'$,$A'$,$F_Z'$ are
defined by minimizing $\chi^2(\beta,A,F_Z,m_\nu)$
in $\beta,A$
and $F_Z$ at fixed $m_\nu$.

Our best fits to the observed data can be seen in 
Fig.~\ref{fit_nu}, for evolution parameter 
$\alpha=3$.
The neutrino mass 
is 
$2.75^{+1.28(3.15)}_{-0.97(1.89)}$~eV 
for the ``halo''-  
and $0.26^{+0.20(0.50)}_{-0.14(0.22)}$~eV
for the ``EG''-case, respectively. The first numbers are
the 1\,$\sigma$, the numbers in the brackets are the
2\,$\sigma$ errors. This gives an absolute lower bound on the mass of the 
heaviest $\nu$ of $0.06$ eV at the 95\% CL. 
Note, that the surprisingly small uncertainties are based on the above
$\chi^2$ analysis and dominantly statistical ones.
The fits are
rather good; for 21 non-vanishing bins and 4 fitted parameters
they can be as low as 
$\chi^2=18.6$. 
We determined $m_\nu$ for a wide range
of cosmological source evolution ($\alpha =0\div 3$) and Hubble 
parameter 
($H_0=(71\pm 7)\times^{1.15}_{0.95}$~km/sec/Mpc) and  
observed only a moderate dependence on them. The results
remain within the above error bars. For these mass scales the atmospheric
or solar $\nu$ experiments suggest practically degenerate $\nu$ masses.
This has no influence on our $\nu$ mass determination, 
but is taken into account in our flux determination. 

We performed a Monte-Carlo analysis studying higher statistics. In the near 
future, Auger~\cite{PAUG} will provide a ten times higher statistics, which 
reduces the error bars in the neutrino mass to $\approx$ one third of their 
present values.  

One of the most attractive patterns for $\nu$ masses is similar to
the one of the charged leptons or quarks: the masses are 
hierarchical, thus the mass difference between the families is 
approximately the mass of the heavier particle. Using the 
mass difference of the atmospheric $\nu$ oscillation for the
heaviest mass~\cite{PDG}, one obtains values between 0.03 and 0.09~eV.  
It is an intriguing feature of our result that the smaller
one of the predicted masses is compatible on the $\approx$ 1.3\,$\sigma$ level 
with this scenario. 

\begin{figure}
\begin{center}
\epsfig{file=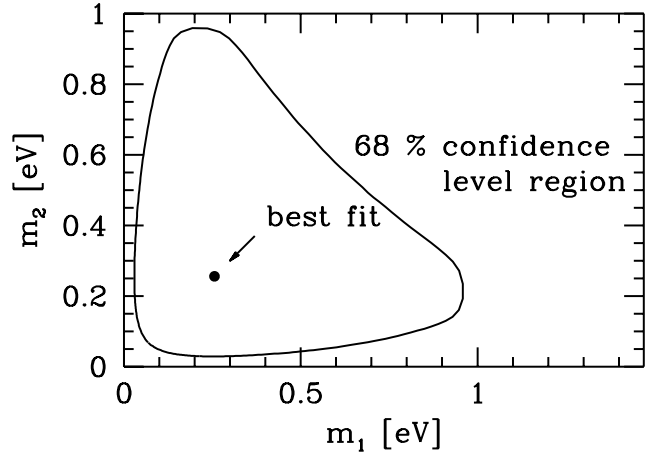,bbllx=20pt,bblly=221pt,bburx=570pt,bbury=608pt,%
width=8.65cm}
\caption{\label{masses}
{
The best fit and the 1\,$\sigma$ (68\% confidence level) region in a scenario
with two non-degenerate $\nu$ masses.
}}
\end{center}
\end{figure}

Another popular possibility is to have 4 neutrino types. Two of them 
-- electron and sterile neutrinos -- are
separated by the solar $\nu$ oscillation solution, the other two
-- muon and tau -- by the atmospheric $\nu$ oscillation 
solution, whereas the mass difference between the two groups is
of the order of 1~eV. We studied this possibility, too. On our
mass scales and resolution the electron and sterile neutrinos are 
practically degenerate with mass $m_1$ and the muon and tau
neutrinos are also degenerate with mass $m_2$. The best fit and the  
1\,$\sigma$ region in the $m_1-m_2$ plane  
is shown in Fig.~\ref{masses} for the ``EG''-case. Since
this two-mass scenario has much less constraints the allowed region 
for the masses is larger than in the one-mass scenario.

\begin{figure}
\begin{center}
\epsfig{file=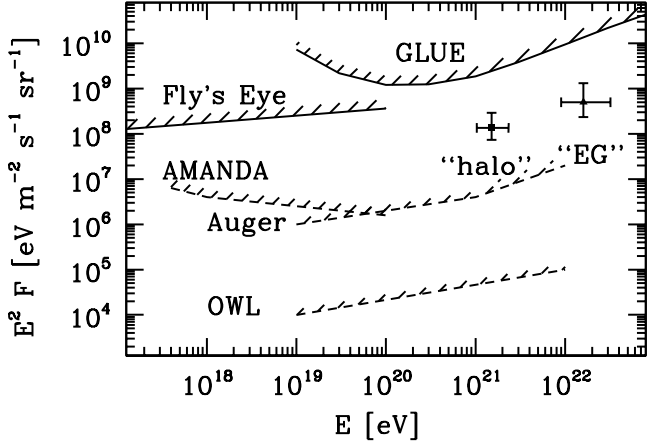,bbllx=20pt,bblly=221pt,bburx=570pt,bbury=608pt,%
width=8.65cm}
\caption[...]{\label{eflux}
Neutrino fluxes, $F = \frac{1}{3}\sum_{i=1}^3 ( F_{\nu_i}+F_{\bar\nu_i})$, 
required by the Z-burst hypothesis for 
the ``halo'' and the ``extragalactic'' case, 
for evolution parameter $\alpha =3$.
The horizontal errors indicate the 1\,$\sigma$ uncertainty of the
mass determination and the vertical errors include also the uncertainty
of the Hubble expansion rate.
The dependence on 
$\alpha$ is just of the order of the thickness of the lines.
Also shown are upper limits from Fly's Eye~\cite{FLY85} and the 
Gold\-stone lunar ultrahigh energy neutrino ex\-pe\-ri\-ment GLUE~\cite{GO01}, 
as well as projected sen\-si\-tivi\-ties of AMAN\-DA~\cite{BA00}, 
Auger~\cite{Y99,CA98} and OWL~\cite{Y99,OR97}.} 
\end{center}
\end{figure}

{\it III. Necessary UHE$\nu$ flux.\,---}
The necessary UHE$\nu$ flux at $E_\nu^{\rm res}$ 
can be obtained via Eqs.~(\ref{nu-flux}) and (\ref{flux}) from our fits. 
We have summarized them in Fig.~\ref{eflux}, together with some
existing upper limits and projected sensitivities of 
present, near future and future observational projects. 
The necessary $\nu$ flux appears to be well below present upper limits
and is within the expected sensitivity of AMANDA, Auger, and OWL.
Clearly, our fluxes are higher than the ones found in Ref.~\cite{Y99} based
on local overdensities $f_\nu$. However, since we also have a background
the normalization of the Z-burst component is different and correspondingly 
our fluxes are somewhat less than a factor of $f_\nu$ higher.   
An important 
constraint for all top-down scenarios~\cite{BS00} is the 
EGRET observation of a diffuse $\gamma$ background~\cite{EGRET}.   
As a cross check, 
we 
calculated the total energy in $\gamma$s from Z-bursts. 
We assumed that all 
energy
ends up between 30 MeV and 100 GeV. 
Our $\gamma$ flux is 
somewhat
smaller than that of EGRET.

{\it IV. Conclusions.\,---}
We compared the predicted spec\-trum of the 
Z-burst hypothesis with the observed
UHE\-CR spectrum. We should emphasize, that
only a realistic 
overdensity of
R$\nu$s was used. 
We determined the
mass of the heaviest R$\nu$:
$m_\nu=2.75^{+1.28}_{-0.97}$ eV 
for halo and
$0.26^{+0.20}_{-0.14}$~eV for EG scenarios. 
The second mass, 
with a lower bound of $0.06$ eV on the 95\% CL, is 
compatible with a 
hierarchical $\nu$ mass scenario with the largest mass suggested by 
the atmospheric $\nu$ oscillation.
The necessary UHE$\nu$
flux
should be detected in the near future.

We thank S. Barwick, 
O. Biebel, S. Bludman, W. Buchm\"uller, 
M. Kachelriess, 
K. Mannheim, H. Meyer, 
W. Ochs, 
G. Sigl, 
and K. Petrovay for useful 
discussions. 
We thank the OPAL collaboration for their unpublished
results on 
Z decays.
This work was partially supported by Hungarian Science Foundation
grants No. OTKA-\-T34980/\-T29803/\-T22929/\-M28413/%
\-IKTA/\-OM-MU-708.

\end{document}